\documentclass[preprint,12pt]{elsarticle}

\newcommand{\di}{\mathrm{i}\mkern1mu}% im i in roman with spacing

\usepackage{bbm}

\usepackage{natbib}
\biboptions{numbers,square,comma,compress}

\usepackage{amssymb}
\usepackage{amsmath}
\journal{Radiation Physics and Chemistry}

\begin{document}

\begin{frontmatter}



%%%%%%%%%%%%%%%%%%%%%%%%%%%%%%%%%%%%%%%%%%%%%%%%%%%%%%%%%%%%%%%%%%%%%%%%

\title{Closed analytical formulas for the Ashley--Ritchie--Brandt\\
and Jackson--McCarthy polarization functions}


\author[1,2]{Jos\'{e} M. Fern\'{a}ndez-Varea}
\ead{jose@fqa.ub.edu}
\cortext[cor1]{Corresponding author}

\author[3]{Alejandro Fern\'{a}ndez-Mrosko}
\ead{mrosko04@gmail.com}

\affiliation[1]{organization={Departament de F\'{\i}sica Qu\`{a}ntica i Astrof\'{\i}sica,Universitat de Barcelona},%Department and Organization
            addressline={Mart\'{\i} i Franqu\`{e}s~1},
            postcode={ES-08028},
            city={Barcelona},
            state={Catalonia},
            country={Spain}}

\affiliation[2]{organization={Institut de Ci\`{e}ncies del Cosmos (ICCUB), Universitat de Barcelona},%Department and Organization
            addressline={Mart\'{\i} i Franqu\`{e}s~1},
            postcode={ES-08028},
            city={Barcelona},
            state={Catalonia},
            country={Spain}}

\affiliation[3]{organization={Facultat de Matem\`{a}tiques i Inform\`{a}tica, Universitat de Barcelona},%Department and Organization
            addressline={Gran Via de les Corts Catalanes~585},
            postcode={ES-08007},
            city={Barcelona},
            state={Catalonia},
            country={Spain}}

%%%%%%%%%%%%%%%%%%%%%%%%%%%%%%%%%%%%%%%%%%%%%%%%%%%%%%%%%%%%%%%%%%%%%%%%

\begin{abstract}

The $Z_{1}^{3}$ correction in the Bethe--Bloch formula that quantifies
the electronic stopping power of matter for heavy charged particles is
revisited. This correction accounts for the Barkas--Andersen effect.
Ashley, Ritchie and Brandt (ARB) wrote it in terms of a non-relativistic
polarization function, and Jackson and McCarthy (JM) obtained the
relativistic counterpart. Both expressions involve intricate triple
integrals. The present manuscript describes how the polarization
functions of JM and ARB can be reduced to compact analytical formulas.
The corresponding asymptotic behaviours are also found. The accuracy of
two currently-used empirical fits to the ARB and JM polarization
functions is assessed.

\end{abstract}


%%%%%%%%%%%%%%%%%%%%%%%%%%%%%%%%%%%%%%%%%%%%%%%%%%%%%%%%%%%%%%%%%%%%%%%%


%\item
%A typo has been fixed in the polarization function of Maynard and Deutsch.

%\item
%An analytical formula has been deduced for the Ashley--Ritchie--Brandt polarization function.

%\item
%An analytical formula has been deduced for the Jackson--McCarthy polarization function.


%%%%%%%%%%%%%%%%%%%%%%%%%%%%%%%%%%%%%%%%%%%%%%%%%%%%%%%%%%%%%%%%%%%%%%%%


%%%%%%%%%%%%%%%%%%%%%%%%%%%%%%%%%%%%%%%%%%%%%%%%%%%%%%%%%%%%%%%%%%%%%%%%

\end{frontmatter}


%%%%%%%%%%%%%%%%%%%%%%%%%%%%%%%%%%%%%%%%%%%%%%%%%%%%%%%%%%%%%%%%%%%%%%%%

\section{Introduction}
\label{section:Introduction}

Swift heavy charged particles (HCPs) like protons, antiprotons, positive
and negative muons, $\alpha$~particles, etc that propagate through a
medium undergo a multitude of inelastic collisions where the target
atoms are excited or ionized. The energy lost in each interaction is
generally small. Hence, the slowing down of HCPs is gradual and may be
described by means of the electronic stopping power, $S_{\mathrm{e}}$.
We do not consider here the nuclear stopping power, which is only
relevant at very low energies.

The 1st-order plane-wave Born approximation yields an expression for the
electronic stopping power where $S_{\mathrm{e}}\propto Z_{1}^{2}$, being
$Z_{1}$ the atomic number of the HCP. The additional assumption that
the velocities of the electrons in the medium are negligible yields the
famous (uncorrected) Bethe formula.

A series of experiments conducted in the 1960s by Barkas and co-workers
\cite{Barkas_etal_PRL_1963} indicated that the electronic stopping power
of a medium for MeV/u negative HCPs is somewhat smaller than that of
positive HCPs with the same $|Z_{1}|$ and speed~$v$. In turn, Andersen
and colleagues \cite{Andersen_etal_NPA_1969} observed that the
electronic stopping power of $\alpha$~particles is more than $4$~times
the value for protons. These findings are known as Barkas--Andersen
effect \cite{Sigmund_2006}. It was quickly understood that it arises
from the 2nd-order Born contribution to $S_{\mathrm{e}}$ which is
proportional to $Z_{1}^{3}$.

In a landmark paper published in 1972, Ashley, Ritchie and Brandt (ARB)
\cite{Ashley_etal_PRB_1972} used classical mechanics to deduce the
contribution of distant collisions to the $Z_{1}^{3}$ correction
modelling the electrons as harmonic oscillators with angular
frequency~$\omega$. Their result also emerges from a quantum mechanical
framework \cite{Hill_Merzbacher_PRA_1974}. The same year, Jackson and
McCarthy (JM) \cite{Jackson_McCarthy_PRB_1972} extended the ARB result
to the relativistic domain. The key quantity in both the ARB and JM
formalisms is the polarization function $\mathcal{I}(\xi)$. The argument
$\xi>0$ combines the dependencies on $\omega$, $v$ and the chosen
minimum impact parameter.

The expressions for the polarization functions obtained by ARB and JM,
$\mathcal{I}_{\mathrm{ARB}}(\xi)$ and $\mathcal{I}_{\mathrm{JM}}(\xi)$
hereafter, involve triple integrals whose numerical evaluation is quite
cumbersome. Therefore, simple empirical fitting formulas are used
invariably for practical calculations. The most popular of them was
proposed in Ref.\ \cite{Basbas_etal_PRA_1978}. It is based on a fit to a
small table of $\mathcal{I}_{\mathrm{ARB}}(\xi)$ values furnished in the
same article. On the other hand, in a very recent comprehensive revision
of the Bethe--Bloch equation for the electronic stopping power of HCPs,
Salvat \cite{Salvat_PRA_2022} developed an empirical formula for
$\mathcal{I}_{\mathrm{JM}}(\xi)$ that fitted a large set of numerical
values computed employing robust integration methods. His fit has been
implemented in the \textsc{sbethe} program \cite{Salvat_Andreo_CPC_2023,
Salvat_Andreo_CPC_2025} that calculates $S_{\mathrm{e}}$ for H and He
ions as well as positive and negative muons and antiprotons.

The authoritative tabulation of electronic stopping powers for H and He
ions included in ICRU Report~49 \cite{ICRU49} was generated using the
Bethe--Bloch equation above $\sim$~$1$--$2$~MeV/u. The $Z_{1}^{3}$
correction was evaluated having recourse to
$\mathcal{I}_{\mathrm{ARB}}(\xi)$ in conjunction with the distribution
of angular frequencies that results from the local-plasma approximation
with the statistical Lenz--Jensen model of the atomic electron density.

The $\mathcal{I}_{\mathrm{ARB}}(\xi)$ function also appears in the
so-called PSS correction to the plane-wave Born cross sections for the
ionization of atomic inner (sub)shells by HCP impact
\cite{Basbas_etal_PRA_1978}. Here too empirical fits are employed, e.g.\
in programs \textsc{isics} \cite{Liu_Cipolla_CPC_1996, Cipolla_CPC_2013}
and \textsc{ercs} \cite{Horvat_CPC_2009, Horvat_CPC_2024}.

In 1982, Maynard and Deutsch (MD) were able to simplify
$\mathcal{I}_{\mathrm{ARB}}(\xi)$ considerably, casting it as an
equation that involves modified Bessel functions of the 1st and 2nd
kinds and only requires a numerical one-dimensional integral
\cite{Maynard_Deutsch_JPL_1982}; we shall denote this formula as
$\mathcal{I}_{\mathrm{ARB}}^{\mathrm{MD}}(\xi)$. MD repeated their
expression in Ref.\ \cite{Deutsch_etal_JPC_1983} and in the recent
review article on ion stopping in dense plasmas
\cite{Deutsch_Maynard_MRE_2016}, but gave only a succinct outline of the
lengthy derivation. Surprisingly (to us), the interesting result
obtained by MD has received very little attention. In fact, a comparison
of $\mathcal{I}_{\mathrm{ARB}}^{\mathrm{MD}}(\xi)$ to the values of
$\mathcal{I}_{\mathrm{ARB}}(\xi)$ tabulated in Ref.\
\cite{Basbas_etal_PRA_1978} revealed large discrepancies, which made us
suspect that the MD formula printed in the aforementioned papers might
contain a typo.

The original (and modest) purpose of the present work was to re-derive
$\mathcal{I}_{\mathrm{ARB}}^{\mathrm{MD}}(\xi)$ so as to identify any
typo that could have creeped into the equations printed in Refs.\
\cite{Maynard_Deutsch_JPL_1982, Deutsch_etal_JPC_1983,
Deutsch_Maynard_MRE_2016}. After achieving this goal, we set out to
evaluate analytically the remaining one-dimensional integral in
$\mathcal{I}_{\mathrm{ARB}}^{\mathrm{MD}}(\xi)$ and thus arrive at
closed analytical formulas for the $\mathcal{I}_{\mathrm{JM}}(\xi)$ and
$\mathcal{I}_{\mathrm{ARB}}(\xi)$ polarization functions. The asymptotic
behaviours of these functions were also deduced.

%%%%%%%%%%%%%%%%%%%%%%%%%%%%%%%%%%%%%%%%%%%%%%%%%%%%%%%%%%%%%%%%%%%%%%%%

\section{Methods}
\label{section:Methods}

The JM polarization function is given by
\cite{Jackson_McCarthy_PRB_1972}
\begin{equation}
\label{def:I_JM}
\mathcal{I}_{\mathrm{JM}}(\xi) =
\mathcal{I}_{1}(\xi) + (1-\beta^{2}) \, \mathcal{I}_{2}(\xi),
\end{equation}
where $\beta=v/c$ ($c$ is the speed of light in vacuum).
In the non-relativistic limit $\beta\to 0$, $\mathcal{I}_{\mathrm{JM}}$
reduces to the ARB polarization function
\cite{Ashley_etal_PRB_1972, Hill_Merzbacher_PRA_1974}
\begin{equation}
\label{def:I_ARB}
\mathcal{I}_{\mathrm{ARB}}(\xi) =
\mathcal{I}_{1}(\xi) + \mathcal{I}_{2}(\xi).
\end{equation}
The two terms that contribute to $\mathcal{I}_{\mathrm{JM}}$ and
$\mathcal{I}_{\mathrm{ARB}}$ are
\begin{align}
\label{def:I_1}
\mathcal{I}_{1}(\xi)
&=
-\int_{\xi}^{\infty} \frac{1}{u^{2}} \, K_{1}(u) \, G_{1}(u) \, \mathrm{d}u,
\\[2mm]
\label{def:I_2}
\mathcal{I}_{2}(\xi)
&=
\int_{\xi}^{\infty} \frac{1}{u^{2}} \, K_{0}(u) \, G_{0}(u) \, \mathrm{d}u,
\end{align}
where $K_{0}$ and $K_{1}$ are modified Bessel functions of the 2nd~kind,
\begin{align}
\label{def:G_1}
G_{1}(u)
&=
\int_{-\infty}^{+\infty}
\frac{\cos(ux)}{(1+x^{2})^{5/2}} \,
%\nonumber
%\\[1mm]
%&\quad
%\times
\Big[ (x^{2}-2)\,F_{1}(u,x) - 3x\,F_{2}(u,x) \Big] \, \mathrm{d}x,
\\[2mm]
\label{def:G_0}
G_{0}(\xi)
&=
\int_{-\infty}^{+\infty}
\frac{\sin(ux)}{(1+x^{2})^{5/2}} \,
%\nonumber
%\\[1mm]
%&\quad
%\times
\Big[ 3x\,F_{1}(u,x) - (1-2x^{2})\,F_{2}(u,x) \Big] \, \mathrm{d}x
\end{align}
and
\begin{align}
\label{def:F_1}
F_{1}(u,x)
&=
\int_{-\infty}^{x} \frac{\sin\!\big[u(x-y)\big]}{(1+y^{2})^{3/2}} \, \mathrm{d}y,
\\[2mm]
\label{def:F_2}
F_{2}(u,x)
&=
\int_{-\infty}^{x} \frac{y\,\sin\!\big[u(x-y)\big]}{(1+y^{2})^{3/2}} \, \mathrm{d}y.
\end{align}

Our goal, to carry out analytically the triple integration in
$\mathcal{I}_{1}$ and $\mathcal{I}_{2}$, turned out to be a demanding
task. The full mathematical derivation is indeed quite lengthy; it is
available from the authors upon reasonable request. Here we limit
ourselves to sketch the main steps of the adopted solution method, which
was inspired by a similar calculation described in Ref.\
\cite{Deutsch_Klarsfeld_PRA_1973}. The starting point was to define the
complex function
\begin{equation}
\label{def:M_5,3}
M_{5,3}(z_{1},z_{2}) =
\int_{-\infty}^{+\infty}
\frac{\mathrm{e}^{\di z_{1}x}}{(1+x^{2})^{5/2}}
\left( \int_{-\infty}^{x} \frac{\mathrm{e}^{-\di z_{2}y}}{(1+y^{2})^{3/2}} \, \mathrm{d}y \right)
\mathrm{d}x.
\end{equation}
Next we wrote $G_{1}$ and $G_{0}$ as linear combinations of
$\operatorname{Im}\!\big(M_{5,3}(z_{1},z_{2})\big)$ and its partial derivatives:
\begin{equation}
\label{G_1}
G_{1}(u) =
-\frac{1}{2}
\left[
\left( 2 + \frac{\partial^{2}}{\partial z_{1}^{2}} + 3 \, \frac{\partial^{2}}{\partial z_{1}\,\partial z_{2}} \right)
\operatorname{Im}\!\big(M_{5,3}(z_{1},z_{2})\big)
\right]^{(z_{1},z_{2})=(2u,u)}_{(z_{1},z_{2})=(0,-u)}
\end{equation}
and
\begin{equation}
\label{G_0}
G_{0}(u) =
-\frac{1}{2}
\left[
\left( 3 \, \frac{\partial}{\partial z_{1}} + \frac{\partial}{\partial z_{2}} + 2 \, \frac{\partial^{3}}{\partial z_{1}^{2}\,\partial z_{2}} \right)
\operatorname{Im}\!\big(M_{5,3}(z_{1},z_{2})\big)
\right]^{(z_{1},z_{2})=(2u,u)}_{(z_{1},z_{2})=(0,-u)}.
\end{equation}
Since $u>0$ in Eqs.\ (\ref{def:I_1}) and (\ref{def:I_2}) only the cases
$z_{1}\geqslant z_{2}\geqslant 0$ and $z_{1}\geqslant 0\geqslant z_{2}$
needed to be considered.
Then, to perform the innermost integrals (over $y$ and $x$) we had to
apply a change of variables, use a liberal dose of complex analysis
involving contour integration along carefully-selected paths in the
complex plane, and repeated integration by parts. All this led to the
result, valid for $z_{1}\geqslant|z_{2}|\geqslant 0$,
\begin{equation}
\label{eq:ImM_5,3}
\begin{split}
\operatorname{Im}\!\big(M_{5,3}(z_{1},z_{2})\big)
&=
\frac{\pi}{24} \left( 3z_{1}^{2}+6z_{1}z_{2}-z_{2}^{2}+3z_{1}+9z_{2} \right) \mathrm{e}^{-(z_{1}-z_{2})}
\\[1mm]
&\quad
- \frac{2\pi}{3} \, z_{1}z_{2} \, \Big[ 2 \, K_{1}(z_{1}) + z_{1} \, K_{0}(z_{1}) \Big] \, I_{1}(z_{2}) \,
\mathbbm{1}_{\{z_{1}\geqslant z_{2}\geqslant 0\}},
\end{split}
\end{equation}
where $I_{1}$ is a modified Bessel function of the 1st kind and
\begin{equation*}
\mathbbm{1}_{\{z_{1}\geqslant z_{2}\geqslant 0\}} \equiv
\begin{cases}
1 & \mbox{if $z_{1}\geqslant z_{2}\geqslant 0$},
\\[1mm]
0 & \mbox{otherwise}.
\end{cases}
\end{equation*}
Finally, the outermost integration (over~$u$) could be done
straightforwardly with the help of the first part of the fundamental
theorem of calculus.

%%%%%%%%%%%%%%%%%%%%%%%%%%%%%%%%%%%%%%%%%%%%%%%%%%%%%%%%%%%%%%%%%%%%%%%%

\section{Results and discussion}
\label{section:ResultsDiscussion}

%=======================================================================

\subsection{Analytical expressions for the polarization functions}

The outcome of the mathematical derivation outlined above are
closed analytical formulas for $\mathcal{I}_{1}$ and $\mathcal{I}_{2}$
in terms of modified Bessel functions of the 1st and 2nd kinds,
\begin{equation}
%\begin{split}
\label{eq:I_1}
\mathcal{I}_{1}(\xi) =
%&=
\pi \,
\bigg\{
- \xi \, K_{1}(2\xi) \, \Big[ K_{0}(\xi) \, I_{0}(\xi) + K_{1}(\xi) \, I_{1}(\xi) \Big]
%\\[1mm]
%&\quad
+ 2\xi \, K_{0}(2\xi) \, K_{1}(\xi) \, I_{0}(\xi)
\bigg\}
%\end{split}
\end{equation}
and
\begin{equation}
\label{eq:I_2}
\mathcal{I}_{2}(\xi) =
\pi \,
\bigg\{
2\xi \, K_{1}(2\xi) \, K_{0}(\xi) \, I_{0}(\xi)
- K_{0}(2\xi)
\bigg\}.
\end{equation}
The sum of $\mathcal{I}_{1}$ and $\mathcal{I}_{2}$ yields a relatively
compact analytical expression for the ARB polarization function,
\begin{equation}
\label{eq:I_ARB}
%\begin{split}
\mathcal{I}_{\mathrm{ARB}}(\xi) =
%&=
\pi \,
\bigg\{
\xi \, K_{1}(2\xi) \Big[ K_{0}(\xi)\,I_{0}(\xi) - K_{1}(\xi)\,I_{1}(\xi) \Big]
%\\[1mm]
%&\quad
+ K_{0}(2\xi) \Big[ 2\xi \, K_{1}(\xi)\,I_{0}(\xi) - 1 \Big]
\bigg\}.
%\end{split}
\end{equation}

For the sake of completeness we also provide the correct expression for
the ARB polarization function reported by MD in Refs.\
\cite{Maynard_Deutsch_JPL_1982, Deutsch_etal_JPC_1983,
Deutsch_Maynard_MRE_2016},
\begin{equation}
%\begin{split}
\mathcal{I}_{\mathrm{ARB}}^{\mathrm{MD}}(\xi) =
%&=
2\pi
\left\{
\xi \, K_{1}(2\xi)
\Big[ K_{0}(\xi)\,I_{0}(\xi) - K_{1}(\xi)\,I_{1}(\xi) \Big]
+ \int_{\xi}^{\infty} \mathrm{d}u \, K_{1}(2u) \, K_{1}(u) \, I_{1}(u)
\right\}.
%\end{split}
\end{equation}
The sought typo happened to be in the argument of the leftmost
$K_{1}$~function.

\begin{figure}
\begin{center}
\includegraphics[height=92mm]{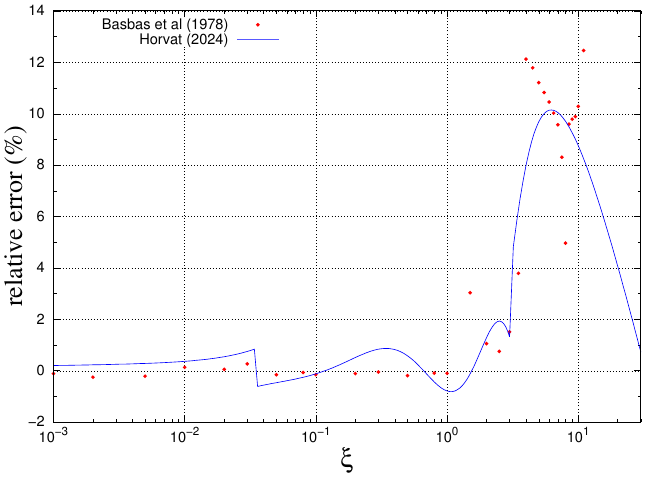}
\caption{
Relative errors (in \%) with respect to the ARB polarization function,
Eq.\ (\ref{eq:I_ARB}).
The red circles correspond to the numerical values given in table~IV of
Ref.\ \cite{Basbas_etal_PRA_1978}.
The blue curve pertains to the empirical fit implemented in the
\textsc{ercs} program \cite{Horvat_CPC_2009, Horvat_CPC_2024}.
}
\label{fig:1}
\end{center}
\end{figure}

Fig.~\ref{fig:1} shows the relative error of the sample numerical values
of $\mathcal{I}_{\mathrm{ARB}}(\xi)$ tabulated in Ref.\
\cite{Basbas_etal_PRA_1978} with respect to our analytical expression,
Eq.\ (\ref{eq:I_ARB}). The deviation is less than $0.5\%$ for
$\xi\lesssim 1$. Then it grows considerably for $\xi\gtrsim 1$ and
reaches $12\%$ at $\xi=11$, suggesting instabilities in the numerical
triple integration. The relative error of the empirical formula employed
by the \textsc{ercs} program \cite{Horvat_CPC_2009, Horvat_CPC_2024} is
also displayed in Fig.~\ref{fig:1}. As expected, it follows the
(erratic) trend of the computed values given in Ref.\
\cite{Basbas_etal_PRA_1978}, The error remains around $1\%$ when
$\xi\lesssim 1$ but for $\xi\gtrsim 1$ it increases to $10\%$.

%.......................................................................
%.......................................................................

\begin{figure}
\begin{center}
\includegraphics[height=92mm]{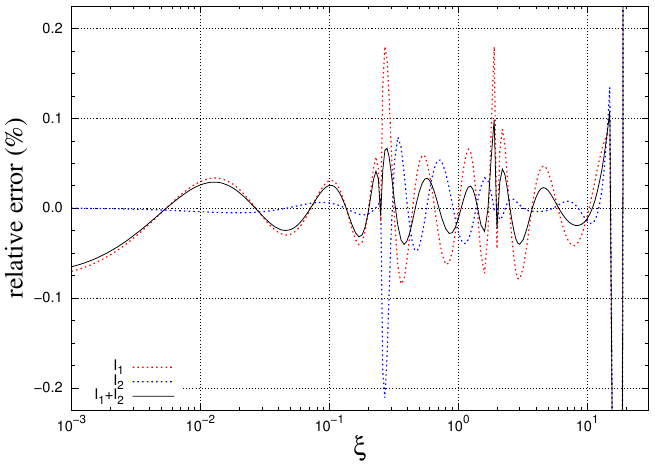}
\caption{
Relative errors (in \%) with respect to the $\mathcal{I}_{1}$,
$\mathcal{I}_{2}$ and $\mathcal{I}_{\mathrm{ARB}}$ functions,
Eqs.\ (\ref{eq:I_1}), (\ref{eq:I_2}) and (\ref{eq:I_ARB}), respectively.
The red and blue dotted curves correspond, respectively, to the
empirical fits for $\mathcal{I}_{1}$ and $\mathcal{I}_{2}$ implemented
in the \textsc{sbethe} program \cite{Salvat_Andreo_CPC_2023,
Salvat_Andreo_CPC_2025}, and the black solid curve is the relative error
of their sum.
}
\label{fig:2}
\end{center}
\end{figure}

The relative errors of the empirical $\mathcal{I}_{1}$ and
$\mathcal{I}_{2}$ functions of Ref.\ \cite{Salvat_PRA_2022},
incorporated in the \textsc{sbethe} program
\cite{Salvat_Andreo_CPC_2023, Salvat_Andreo_CPC_2025}, are plotted in
Fig.~\ref{fig:2}. They do not exceed $0.2\%$ for $0<\xi\lesssim 15$.
And, as stated in Refs.\ \cite{Salvat_PRA_2022, Salvat_Andreo_CPC_2023},
the deviation of $\mathcal{I}_{1}+\mathcal{I}_{2}$ (i.e.\
$\mathcal{I}_{\mathrm{ARB}}$) is at most $0.1\%$ in the same interval.

%.......................................................................
%.......................................................................

%=======================================================================

\subsection{Aymptotic behaviours of the polarization functions}

The functions $\mathcal{I}_{1}$, $\mathcal{I}_{2}$ and
$\mathcal{I}_{\mathrm{ARB}}$ display the following asymptotic behaviours
when $\xi\to\infty$:
\begin{align}
\label{eq:I_1_asymptotic_infty}
\mathcal{I}_{1}(\xi)
&\underset{\xi\to\infty}{\sim}
\frac{1}{8} \left(\frac{\pi}{\xi}\right)^{3/2} \mathrm{e}^{-2\xi}
+ \mathcal{O}\!\left(\frac{\mathrm{e}^{-2\xi}}{\xi^{5/2}}\right),
\\[2mm]
\label{eq:I_2_asymptotic_infty}
\mathcal{I}_{2}(\xi)
&\underset{\xi\to\infty}{\sim}
\frac{1}{8} \left(\frac{\pi}{\xi}\right)^{3/2} \mathrm{e}^{-2\xi}
+ \mathcal{O}\!\left(\frac{\mathrm{e}^{-2\xi}}{\xi^{5/2}}\right),
\\[2mm]
\label{eq:I_ARB_asymptotic_infty}
\mathcal{I}_{\mathrm{ARB}}(\xi)
&\underset{\xi\to\infty}{\sim}
\frac{1}{4} \left(\frac{\pi}{\xi}\right)^{3/2} \mathrm{e}^{-2\xi}
+ \mathcal{O}\!\left(\frac{\mathrm{e}^{-2\xi}}{\xi^{5/2}}\right).
\end{align}
JM predicted an exponential decrease of $\mathcal{I}_{1}$ and
$\mathcal{I}_{2}$ for large values of the argument but did not give
explicit formulas \cite{Jackson_McCarthy_PRB_1972}.
On the other hand, in the case of $\mathcal{I}_{\mathrm{ARB}}$ MD
arrived at Eq.\ (\ref{eq:I_ARB_asymptotic_infty})
\cite{Maynard_Deutsch_JPL_1982, Deutsch_Maynard_MRE_2016}.

In turn, the corresponding behaviours for $\xi\to 0$ ($\xi>0$) are
\begin{align}
\label{eq:I_1_asymptotic_0}
\mathcal{I}_{1}(\xi)
&\underset{\xi\to 0}{\sim}
- \frac{3\pi}{2} \ln\xi
- \frac{\pi}{4} \left( 6\gamma + 2\ln 2 + 1 \right)
- \frac{23\pi}{8} \, \xi^{2} \left[ \ln\xi + \gamma - \frac{3\ln 2}{23} - \frac{3}{4} \right]
\nonumber
\\[1mm]
&\quad
+ \mathcal{O}(\xi^{4}\ln^{2}\xi),
\\[2mm]
\label{eq:I_2_asymptotic_0}
\mathcal{I}_{2}(\xi)
&\underset{\xi\to 0}{\sim}
\pi\ln 2
%\nonumber
%\\[1mm]
%&\quad
- 2\pi \xi^{2}
\left[
\ln^{2}\xi
+ \left( 2\gamma - \ln 2 - \frac{3}{4} \right) \ln\xi
+ \gamma^{2}
- \left( \ln 2 + \frac{3}{4} \right) \gamma
+ \frac{\ln 2}{4}
+ \frac{3}{8}
\right]
\nonumber
\\[1mm]
&\quad
+ \mathcal{O}(\xi^{3}\ln^{2}\xi),
\\[2mm]
\label{eq:I_asymptotic_0}
\mathcal{I}_{\mathrm{ARB}}(\xi)
&\underset{\xi\to 0}{\sim}
- \frac{3\pi}{2} \ln\xi
- \frac{\pi}{4} \left( 6\gamma - 2\ln 2 + 1 \right)
\nonumber
\\[1mm]
&\quad
- 2\pi \xi^{2}
\left[
\ln^{2}\xi
+ \left( 2\gamma - \ln 2 + \frac{11}{16} \right) \ln\xi
+ \gamma^{2}
- \left( \ln 2 - \frac{11}{16} \right) \gamma
+ \frac{\ln 2}{16} - \frac{45}{64}
\right]
\nonumber
\\[1mm]
&\quad
+ \mathcal{O}(\xi^{3}\ln^{2}\xi),
\end{align}
where $\gamma$~denotes the Euler--Mascheroni constant. According to JM,
numerically $\mathcal{I}_{1}(\xi)\approx -\frac{3\pi}{2}\ln(8\xi/3)$ and
$\mathcal{I}_{2}(\xi)\approx 2.175$ for $\xi\to 0$
\cite{Jackson_McCarthy_PRB_1972}. The former approximation is inaccurate
whereas the latter is close to the exact value $\pi\ln 2$. In the case
of $\mathcal{I}_{\mathrm{ARB}}$, Eq.\ (\ref{eq:I_asymptotic_0}) agrees
with the behaviour already found by MD. However, these authors quoted
the constants as numbers with just two or three decimals
\cite{Maynard_Deutsch_JPL_1982, Deutsch_Maynard_MRE_2016}; for instance,
instead of $2\gamma-\ln 2+\frac{11}{16}$ they wrote $1.14$.

%%%%%%%%%%%%%%%%%%%%%%%%%%%%%%%%%%%%%%%%%%%%%%%%%%%%%%%%%%%%%%%%%%%%%%%%

\section{Conclusions}
\label{section:Conclusions}

We have managed to reduce the convoluted original expressions for the JM
and ARB polarization functions $\mathcal{I}_{\mathrm{JM}}(\xi)$ and
$\mathcal{I}_{\mathrm{ARB}}(\xi)$ to closed analytical formulas. The
former can be easily incorporated in programs like \textsc{sbethe}
\cite{Salvat_Andreo_CPC_2023, Salvat_Andreo_CPC_2025} to compute the
Barkas--Andersen correction term in the Bethe--Bloch equation. The
latter may be included in programs such as \textsc{isics}
\cite{Liu_Cipolla_CPC_1996, Cipolla_CPC_2013} and \textsc{ercs}
\cite{Horvat_CPC_2009, Horvat_CPC_2024} that calculate inner-shell
ionization cross sections by the impact of HCPs, replacing the empirical
fits that they currently implement. On the other hand, the low-$\xi$ and
high-$\xi$ limits might perhaps be useful to devise accurate empirical
fits if nevertheless very simple formulas for
$\mathcal{I}_{\mathrm{ARB}}(\xi)$ and $\mathcal{I}_{\mathrm{JM}}(\xi)$
are preferred.

%%%%%%%%%%%%%%%%%%%%%%%%%%%%%%%%%%%%%%%%%%%%%%%%%%%%%%%%%%%%%%%%%%%%%%%%

\section*{Declaration of competing interest}

The authors declare that they have no known competing financial
interests or personal relationships that could have appeared to
influence the work reported in this paper.

%%%%%%%%%%%%%%%%%%%%%%%%%%%%%%%%%%%%%%%%%%%%%%%%%%%%%%%%%%%%%%%%%%%%%%%%

\section*{Acknowledgements}

The work of J~M~F-V is supported by the Spanish MICIU/AEI/10.13039/501100011033
through grant PID2023-147112NB-C22.

%%%%%%%%%%%%%%%%%%%%%%%%%%%%%%%%%%%%%%%%%%%%%%%%%%%%%%%%%%%%%%%%%%%%%%%%

%% Labels are used to cross-reference an item using \ref command.

%\appendix
%\section{Example Appendix Section}
%\label{app1}


%%%%%%%%%%%%%%%%%%%%%%%%%%%%%%%%%%%%%%%%%%%%%%%%%%%%%%%%%%%%%%%%%%%%%%%%

%%       \cite{<label>} ==> [1]

%%
%Example citation, See \cite{lamport94}.


\bibliographystyle{elsarticle-num}
\bibliography{polarization_function.bib}



%\begin{thebibliography}{00}

%% For numbered reference style

%\bibitem{lamport94}
%  \textit{\LaTeX: a document preparation system},
%  Addison Wesley, Massachusetts,

%\end{thebibliography}

%%%%%%%%%%%%%%%%%%%%%%%%%%%%%%%%%%%%%%%%%%%%%%%%%%%%%%%%%%%%%%%%%%%%%%%%

\end{document}